# Spin negative differential resistance in edge doped zigzag graphene nanoribbons

*Chao Jiang[1], Xue-Feng Wang[1,2*], Ming-Xing Zhai[1]*

[1]School of Physical Science and Technology, Soochow University, Suzhou, China 215006

[2]Shanghai Institute of Microsystem and Information Technology, Chinese Academy of Sciences,

865 Changning Road, Shanghai 200050 China

**abstract**

The nonlinear spin-dependent transport properties in zigzag graphene nanoribbons (ZGNRs) edge doped by an atom of group III and V elements are studied systematically using density functional theory combined with non-equilibrium Green's functions. The dopant type, acceptor or donor, and the geometrical symmetry, odd or even, are found critical in determining the spin polarization of the current and the current-voltage characteristics. For ZGNRs substitutionally doped on the lower-side edge, the down (up) spin current dominates in odd-(even-)width ZGNRs under a bias voltage around 1V. Remarkably, in even-width ZGNRs, doped by group III elements (B and Al), negative differential resistance (NDR) occurs only for down spins. The bias range of the spin NDR increases with the width of ZGNRs. The clear spin NDR is not observed in any odd-width ZGNRs nor in even-width ZGNRs doped by group V elements (N, and P). This peculiar spin NDR of edge doped ZGNRs suggests potential applications in spintronics.

* Corresponding author: Tel/Fax: +86 512 65226297
  E-mail address: xf_wang1969@yahoo.com (X.-F Wang)

## 1. Introduction

Graphene nanoribbons (GNRs), the one-dimensional strips of graphene, are promising materials for quantum electronic devices due to their atomistic size [1,2], peculiar electronic structure [3-7], and outstanding transport properties [3,8-15]. Based on the crystallographic orientation of their edges, GNRs are classified into two primary categories: the zigzag graphene nanoribbons (ZGNRs) and the armchair graphene nanoribbons (AGNRs) [16,17]. In case that the spin effect is disregarded, ZGNRs are expected to be metallic whereas AGNRs can be metallic or semiconducting depending on their width. The electronic structure of AGNRs can be approximately obtained from that of graphene, but it is not true for ZGNRs due to the appearance of edge states therein [5]. Remarkably, resulting from the dangling $p_z$ orbitals of the edge carbon atoms, two flat energy bands corresponding to edge states may appear at the Fermi energy in the $k$ range $[2\pi/3a, \pi/a]$. Here $a = \sqrt{3}a_0$ is the lattice constant of the ZGNRs with $a_0$ the bond length between carbon atoms [18-20]. This electronic structure with high density of states (DOS) at the Fermi energy is not stable and the electron-electron interaction introduces an energy gap at the Fermi energy which can be described by the Hubbard model [16,20,21]. The system becomes locally spin polarized with an antiferromagnetic ground state: the atoms on the lower (upper) edge are associated to sublattice A (B) and are up- (down-) spin polarized.

The appearance of magnetism in ZGNRs composed of nonmagnetic elements suggests potential applica-tions in spintronics [22]. Nevertheless, the intrinsic ZGNR system as a whole is not magnetic due to its symmetry. Various ways to break the symmetry have been proposed, e.g. using a substrate to introduce a potential difference between the A and B sublattices [7]; applying an electric potential difference between the two edges [23,24]; introducing geometric disorders [25-27], defects [17], or impurities [28-31] to the system. It is well known that the substitutional doping of atoms in groups III and V is the most employed technique in silicon-based semiconductor devices. Similarly, substitutional doping of elements in groups III and V has attracted much attention for manipulating electronic properties of graphene materials in the past few years [23,32-42]. Different from conventional bulk silicon systems, the ZGNRs are on an atomic scale and the edges play important role. It is easier to substitutionally dope an N atom on the edge sites than on other sites [33] and the group III (V) impurities may change from p-type to n-type (n-type to p-type) when being moved from center sites to edge sites [32,33,36] as a result of the competition between the Coulomb interaction and correlation [32].

The phenomenon of negative differential resistance (NDR) is the mechanism for many essential electronic devices including high-frequency oscillators, frequency multipliers, memories, and fast switches. It is then very interesting to realize NDR for graphene based devices [43]. The NDR appears under some conditions and may be manipulated by strain, chirality, defects, and oxidization in GNRs [44-47]. Usually the NDR appears for both up and down spins in the same bias regime [47], but recently we have observed strongly spin-dependent NDR effects in Be-doped, even-width ZGNRs [48].

In this paper, we study nonlinear transport through ZGNRs substitutionally doped on one edge by a single atom of the most favorable doping elements in groups III (B and Al) and V (N and P). In even-width ZGNRs doped by a group-III atom on the lower edge, the NDR phenomenon appears in the current-voltage curves of only one spin orientation at relatively low bias voltage. This unique phenomenon is expected to be very useful for spintronics.

## 2. Systems and Computational Methods

To simulate the current-voltage characteristics of ZGNRs, we consider a device system as illustrated in Figure 1a. An $n$-ZGNR of width $n$ is partitioned into three parts: the semi-infinite left electrode (L), the central scattering region (C), and the semi-infinite right electrode (R). When a bias voltage $V_b$ is applied between the two electrodes, we assume that the Fermi energies in the electrodes are constant as can be realized in

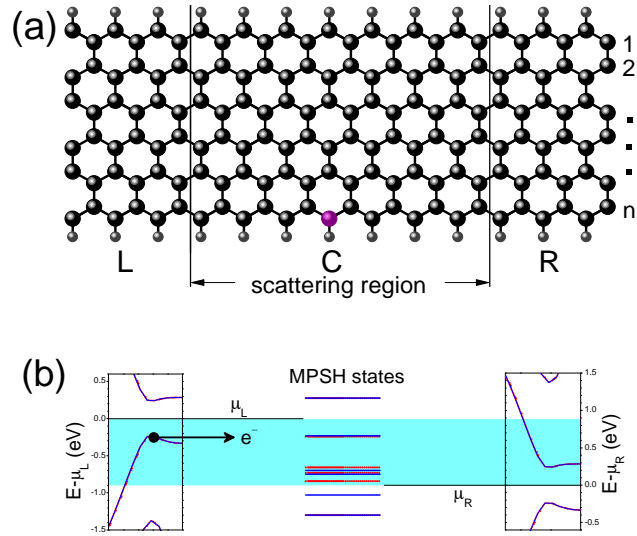

Figure 1. (a) Atomistic structure of an *n*-ZGNR two-probe junction with the width number *n*=6. The position of the C (big dots) and H (small dots) atoms are indicated in black and the doping atom in purple. (b) Energetic scheme of the electron transport model. The scattering region is described by the molecule-projected self-consistent Hamiltonian (MPSH) energy states and the electrodes by the energy bands.

experiments [9,10,43]. The dangling bond of each edge C atom is terminated by an H

atom. The length of the central scattering region is chosen long enough (7 primitive cells here) so that no direct tunneling occurs between the electrodes and the effect of the possible dopant on the electrodes is negligible for the transport study. In the pristine system, the ground state is antiferromagnetic and the edge states of opposite spins are almost equally located away from the chemical potential in energy [21]. At the bottom (top) of the conduction (valence) band, the upper (lower) edge state is spin-up polarized while the lower (upper) edge state is spin-down polarized. We assume that the spin polarizations of the two electrodes are parallel to each other on either edge. In doped systems, the central C atom on the lower edge of the scattering region is substituted by a dopant atom. The energy-band view of the system is shown schematically in Figure 1b when a bias $V_b = (\mu_L - \mu_R)/e$ is applied between the electrodes. The scattering region can be modeled by the eigenstates of the MPSH. In this picture, an electron can tunnel from the energy bands in electrode L to those in electrode R through the MPSH states.

The transport simulation is performed by the atomistix toolkits (ATK) [49,50] in which the quantum transport method of the self-consistent nonequilibrium Green's functions (NEGF) and the *ab initio* density-functional theory (DFT) are implemented. Before the transport computation, the structures are optimized by the ATK and the Vienna ab initio simulation package (VASP) until the atomic forces are less than 0.02 eV/Å. The exchange correlation potential resorts to the spin-polarized local-density approximation (SLDA). The basis set of Single Zeta Polarization (SZP) is used in the calculation since it is good enough in our cases [47,48]. The fineness of the real-space grid is determined by an equivalent plane-wave cutoff of 150Ry; the mesh grid of the k space is 1×1×500 and the electronic temperature is 300 K.

## 3. Results and Discussions

We present the current-voltage ($I-V$) curves of some B, Al, N, and P-doped ZGNRs in Figure 2. The $I-V$ curves show the semiconductor characteristic with a threshold voltage around 0.4V. In the left column of panel (a) for the B-doped 3,5,7-ZGNRs, both spin-up and spin-down currents increase with the bias voltage except a small decrease.

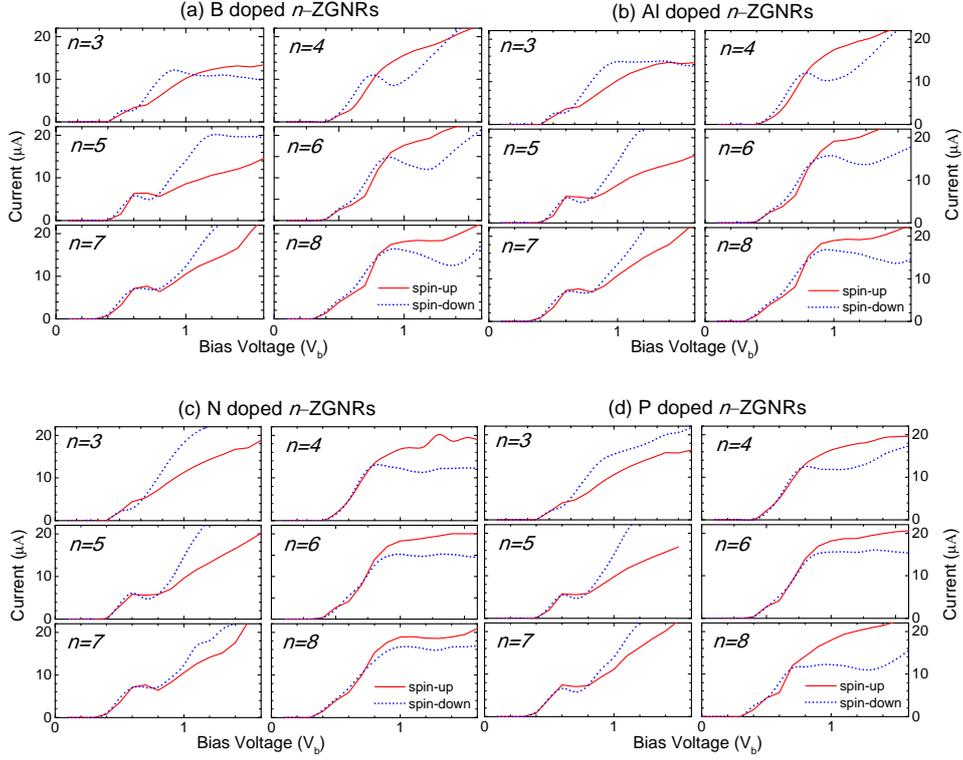

Figure 2. The $I-V$ curves of $n$-ZGNRs, $n=3,4,5,6,7,8$, edge doped by a single (a) B atom, (b) Al atom, (c) N atom, and (d) P atom are presented. The red solid and blue dotted curves are for spin up and spin down electrons, respectively.

The $I-V$ curve for the Al, N, and P-doped 3,5,7-ZGNRs show similar characteristic as illustrated in panel (b), (c), and (d), respectively. Interestingly, a distinguished behavior is observed in 4,6,8-ZGNRs whose structures are symmetric with respect to its central line. For example, in the B-doped 6-ZGNR as shown in the right column of panel (a), the spin-up current increases with the bias monotonously above the threshold. In contrast, the spin-down current reaches a maximum at about 0.9V and then decreases significantly to a minimum at around 1.2V before increases again. In other words, the NDR occurs between 0.9-1.2V *only for spin-down electrons.* This spin NDR is also obviously observed in other B-doped and Al-doped even *n*-ZGNRs as shown in the right column of panels (a) and (b), respectively. The NDR range enlarges and shifts to higher voltage with the width of ZGNRs. The corresponding peak-to-valley ratios are 1.23 and 1.14 for the B and Al doped 6-ZGNRs, respectively, and vary with the width number *n*. In the N and P-doped even *n*-ZGNRs as shown in the right column of panels (c) and (d), respectively, the spin-up current increases with the bias similar to that in the B-doped ones. However, the spin-down current increases and saturates. Our simulation of the $I-V$ curves in *n*-ZGNRs of various width numbers *n* suggests common characteristics in ZGNRs of the same symmetry and with the same type of dopant. The threshold voltage decreases with the ribbon width as a result of the narrowing of the energy gap between the valence and the conduction bands [4]. In ZGNRs of odd width, a current step appears near 0.6V for both the spin-up and spin-down $I-V$ curves. In even-width ZGNRs, the spin-up current increases monotonically with the bias but the spin-down current has a NDR region if a group III atom is doped on the lower edge and saturates at high bias voltage if a group V atom is doped. The spin NDR appears obviously only in even *n*-ZGNRs edge doped by group III elements but not in odd *n*-ZGNRs. Its range extends with the width of the ribbons. Overall, under high bias voltage the spin-down current dominates in odd edge doped ZGNRs but spin-up current dominates in even edge doped ZGNRs.

To understand the underlying mechanism of the spin NDR observed in even ZGNRs edge doped by a group III atom, we present the transmission spectra of 6-ZGNRs doped by a B or N atom in Figure 3. At small bias $V_b = 0.3$V, the energy band gaps of the two electrodes overlap and a wide transmission gap exists in the energy range [−0.35, 0.35] eV for both spin-up and spin-down electrons. In addition, at the energies of the impurity states [35,36], a transmission dip above the transmission gap appears (at 0.55eV for spin-up and at 0.65eV for spin-down electrons) in the B-doped ZGNRs and one below the gap (at −0.75eV for spin-up and at −0.65eV for spin-down electrons) in N-doped ZGNRs. The transmission spectra are highly spin-dependent only near the energies of the impurity states. Since the transport window between the chemical potentials in the electrodes [−e$V_b$/2, e$V_b$/2] is inside the transmission gap, the current is almost zero.

As the bias increases to $V_b > 0.4$V, the band gaps in the two electrodes mismatch and electrons in the valence band of electrode L may tunnel through the scattering region to the conduction band of electrode R. The device shows bipolar characteristics and the transmission gap splits. A transmission structure of width (e$V_b$ − 0.4eV) appears inside the transport windows and the current through the device is proportional to the area of the structure for each spin. The shape of the structure depends on the energy-dependent overlap between wave functions in electrode L or R and in the scattering region. At $V_b = 0.6$ V, the transmission structures of the B-doped and N-doped 6-ZGNRs are similar and both have a higher spin-down component.

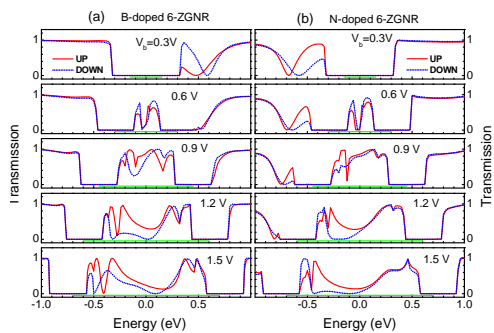

Figure 3. Spin-up (read solid) and spin-down (blue dotted) transmission spectra of electrons are presented for (a) B-doped 6-ZGNR, and (b) N-doped 6-ZGNR. Results under bias $V_b$ =0.3, 0.6, 0.9, 1.2, and 1.5V are shown in panels from top to bottom, respectively. The horizontal green bars above the *x*-axis indicate the transport windows.

At $V_b = 0.9$V, the transmission structures widens and the spin-up components begin to dominate at some energies. When the bias increases further, the structures split into two parts and form a slowly varying region between them. The high-energy part is not much spin-dependent while the low-energy part is strongly spin-dependent. In B-doped 6-ZGNRs, the spin-down component of the low-energy part decreases quickly as $V_b > 0.9$V and reaches a wide gap at $V_b = 1.2$V before recovering again at higher bias voltage. This explains the NDR that occurs for spin-down current in the B-doped 6-ZGNR. The spin-down component of the low-energy part in the N-doped 6-ZGNRs decreases also but not as quickly as in B-doped 6-ZGNRs, resulting in the observed current saturation instead of an obvious NDR region.

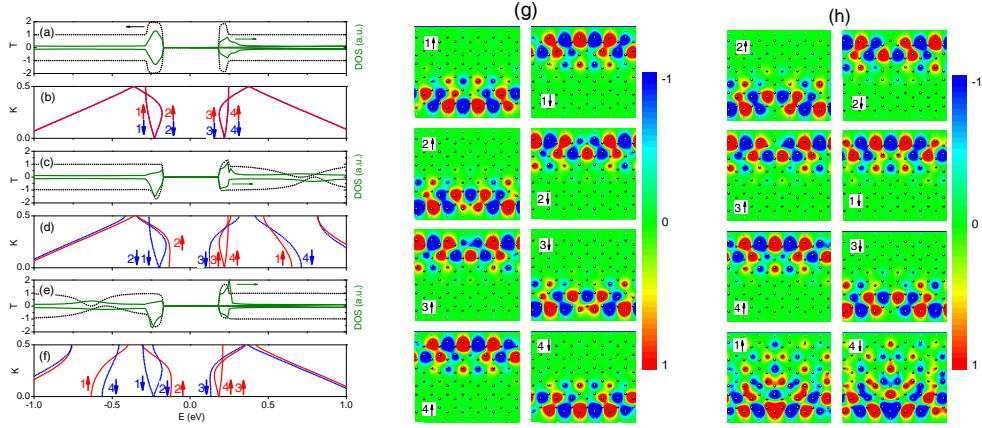

Figure 4. The transmission spectra (solid) and DOS (dashed) of a pristine, B-doped, and N-doped 6-ZGNR two-probe junctions are plotted in (a), (c), and (e), respectively. The positive (negative) parts are for spin-up (down) values. The energy bands (solid for spin-up and dashed for spin-down) of the virtual bulks in which the unit cells are the central regions of the above junctions are presented in (b), (d), and (f), respectively. The real-space wave functions (in a plane slightly above the ZGNR plane) of the k=0 states in bands 1-4 are plotted in (g) for a pristine ZGNR and (h) for a B-doped ZGNR, where the black dots marks the position of the atoms in the scattering region. The up and down arrows indicate the up and down spin polarizations, respectively. The chemical potential is set equal to zero.

From the bias dependence of the transmission spectra we can see that the appearance of the spin NDR in B-doped 6-ZGNRs is closely related to the variation of the electron states in the doped scattering region under the bias. To see what happens in detail when an impurity atom is doped in this region with the effect of the electrodes taken into account, we establish a virtual bulk with its unit cell the same as the scattering region. We calculate the energy bands of the virtual bulk and analyze how the impurities affect the energy bands. In Figure 4, we compare the B-doped and the N-doped 6-ZGNRs with the pristine 6-ZGNRs for their band structures of the virtual bulk as well as the transmission and the DOS of the corresponding two-probe systems.

In a two-probe junction of pristine 6-ZGNRs, as shown in Figure 4a, the transmission is 100% for both spin-up and spin-down electrons in the band of the edge states except near the band edge where the band structure bends (see Figure 1). The energy bands of the corresponding virtual bulk plotted in Figure 4b are the folded bands of a pristine 6-ZGNRs bulk. The four bands besides the energy gap, denoted as 1-4, correspond to the four energy states near the Fermi energy in the scattering region. The spin-up and spin-down states are degenerate but their wave functions are separated in real space as illustrated in Figure 4g. For states 1 and 2 in the valence bands, the spin-up component is confined in the lower edge of the ribbon while the spin-down part in the upper edge. State 1 (state 2) has an antinode (a node) in the middle of the region on the edge. This phase difference between the two states may show its importance and lead to energy separation in edge-doped ribbons. The real space distribution of states 3 and 4 in the conduction bands is the same as that of states 2 and 1, respectively, with opposite spins.

When a B atom is doped in the middle of the scattering region on the lower edge, a transmission dip appears in the conduction band around 0.75eV above the chemical potential [35,36] as shown in Figure 4c. This is attributed to the appearance of impurity states at this energy as indicated by the DOS curve in Figure 4c. The B atom works as a donor impurity and losses 0.3e charge mainly to the carbon atoms on the other edge [35,36]. Our simulation shows that the impurity states are shifted spin-up state 1 and spin-down state 4 as shown in Figure 4d. These two states have an antinode at the position of the doping atom and are affected greatly by the impurity. With the energy increase, the two states mix with the bulk states of higher energies and become slightly extended as shown in Figure 4h. Note here that down-spin states 1 and 2 exchange their position in energy sequence.

In contrast, when an N atom is doped at the same position, spin-up state 1 and spin-down state 4 also shift but to an opposite direction as illustrated in Figure 4e. The N atom works as an acceptor impurity and attracted about 0.3e charge from the carbon atoms around [35,36]. As shown in Figure 4d, the impurity states lead to DOS peaks and transmission dips around 0.65eV below the chemical potential in the two-probe systems. The above analysis is confirmed by the following calculation of the eigenstates of the MPSH in the scattering region.

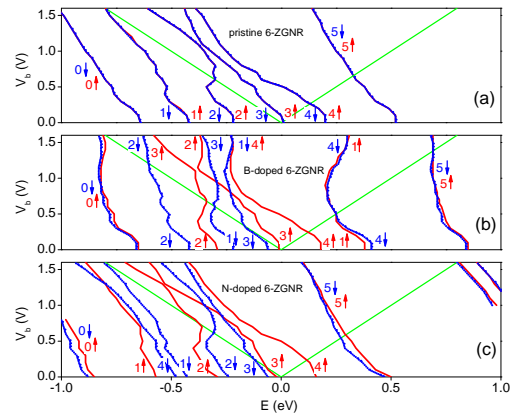

Figure 5. The energy variation of the MPSH eigenstates in the scattering region under applied bias voltage $V_b$ for (a) pristine, (b) B-doped, and (c) N-doped 6-ZGNRs. The edges of the transport windows are marked by the green lines.

In Figure 5a we plot the energies of the MPSH eigenstates versus bias $V_b$ for the pristine 6-ZGNR junction [47]. Comparison of the real-space wave functions between these MPSH states and the $k=0$ states 1-4 in the corresponding virtual bulk system, as described in Figure 4, indicates that states 1-4 correspond to the second highest occupied molecular orbital (HOMO−1), the highest occupied molecular orbital (HOMO), the lowest unoccupied molecular orbital (LUMO), and the second LUMO (LUMO+1) states of the MPSH, respectively. With the increase of $V_b$, all the MPSH energy levels shift to lower energy due to the variation of the electrostatic potential profile and the charge redistribution. Levels 2 and 3 anticross at around $V_b$ =1V. When the two energy levels anticross, their wave functions mix with each other and the real-space distributions of the eigenstates are greatly modified. This may affect significantly the coupling between states in the electrodes and the scattering region and then the I-V characteristics. The NDR may appear in this bias range [47].

In the B-doped case, the symmetry of the system is broken and the up- and down-spin electrons feel a different electrostatic potential profile. Having antinode at the position of the B atom, spin-up state 1 and spin-down state 4 shift significantly upward and approach the energy of the third LUMO (LUMO+2) state 5. They interact and mix with state 5 and become unconfined to the lower edge. Spin-down state 1 has an energy higher than the spin-down state 2. The spin degeneracy is then lifted and the scattering region becomes magnetic with down-spin the majority spin polarization as shown in Figure 5b. The above characteristics are consistent with the virtual bulk analysis described in Figure 4 and explain the strong spin-dependence of the electron transport through the two-probe system. When a bias is applied, unlike in the pristine and the N-doped 6-ZGNRs where all the MPSH levels move down quickly with the bias, the MPSH levels in B-doped 6-ZGNR may shift upward at high bias. Spin-up level 3 moves downward quickly and anticrosses with spin-up state 2 at bias around 1.1V. We can see that the minimal energy separation between the two levels around the anticross region is very small. This indicates that the coupling between the two states is small. In contrast, the spin-down level 1 anticrosses with the spin-down level 3 in the range [0.8, 1.2] V with a much bigger energy separation. This suggests that the strong coupling between the down-spin states 1 and 3, when they anticross, greatly modifies their wave functions. The variation of the wave functions may change their coupling to the states in the electrodes and lead to the conductance decrease in this bias and in the energy regime as shown in Figure 3c. As a result, the spin-down current decreases and the obvious spin NDR takes place. At high bias $V_b$ =1.6V, beyond the anticrossing regime, the wave functions recover their initial patterns at zero bias. In addition, the energy sequence of the MPSH states 1-4 becomes highly correlated with its real-space distribution. This suggests that the electrostatic energy begins to play an important role in determining the energy of the states at high bias.

In the N-doped case, the impurity atom works as an acceptor. The MPSH spin-up level 1 and spin-down level 4 move downward instead of upward and approach the third HOMO (HOMO-2) state 0, as shown in Figure 5c, in agreement with the virtual bulk analysis. The anticrossing between spin-down level 2 and spin-down level 3 occurs in a higher and wider bias range [0.9, 1.4 V] with weaker coupling compared to the B-doped case. The current saturates instead of decreasing correspondingly. Usually more spin-up MPSH levels are located inside the transport windows in doped 6-ZGNRs and this explains why the spin-up current is usually bigger than the spin-down one under high bias.

## 4. Conclusions

We have simulated the spin-dependent current-voltage characteristics in edge doped ZGNRs of various widths employing the density functional theory combined with the Green's function method. A spin NDR is predicted in even $n$-ZGNRs edge doped by an atom of group III elements such as B and Al. In edge doped $n$-ZGNRs where the down-spin is assumed the majority spin, the spin-down current dominates for odd $n$ and the spin-up current dominates for even $n$ under high bias. The impurity atom of group III (V) elements works as a charge donor (acceptor).

**Acknowledgements**

X.-F Wang appreciates P Vasilopoulos for helpful discussion. This work was supported by the National Natural Science Foundation of China (Grant Nos 11074182 and 91121021). This is a Project Funded by the Priority Academic Program Development of Jiangsu Higher Education Institutions.